\documentclass[journal]{IEEEtran}
\usepackage{cite}
\ifCLASSINFOpdf
\else
\fi
\usepackage{float}
 \pdfoutput=1 
\usepackage{amsmath}
\usepackage{amsthm}
\usepackage{amssymb}
\usepackage{array}
\usepackage{url}
\usepackage{graphicx}
\usepackage{caption}
\usepackage{xfrac}
\usepackage{color,soul}
\usepackage{authblk}
\usepackage{graphicx}
\usepackage{booktabs}
\usepackage{multirow}
\usepackage{flushend}

\usepackage{soul,xcolor}

\pdfoutput=1

\begin{document}
	\title{An Experimental Study of NOMA \\ for Connected Autonomous Vehicles}
     \author[$*$]{Eray~Güven}
     \author[$*$]{Caner Göztepe}
     \author[$*$]{Mehmet~Akif~Durmaz,~\IEEEmembership{Member,~IEEE,}}
      \author[$*$]{Semiha~Tedik~Basaran,~\IEEEmembership{Member,~IEEE,}}
      \author[$*$]{Gunes~Karabulut~Kurt,~\IEEEmembership{Senior Member,~IEEE}}
     \author[$\S$]{Oğuz~Kucur}
 


\author{Eray~Güven,
        Caner Göztepe,
        Mehmet~Akif~Durmaz,~\IEEEmembership{Member,~IEEE,}
        Semiha~Tedik~Başaran,~\IEEEmembership{Member,~IEEE,}        Güneş~Karabulut~Kurt,~\IEEEmembership{Senior Member,~IEEE}
        and~Oğuz~Kucur
        
 \thanks{E. Güven and G. Karabulut Kurt are with the {Department of Electrical Engineering,  Polytechnique Montr\'eal, Montr\'eal, Canada, e-mails: \{guven.eray, gunes.kurt\}@polymtl.ca}.}
 \thanks{C. Göztepe, M. A. Durmaz, S. Tedik Başaran are with the {Department of Electronics and Communication Engineering,  Istanbul Technical University, Istanbul, Turkey, e-mails: \{goztepe, durmazm, tedik\}@itu.edu.tr}.}
 \thanks{O. Kucur is with the Department of Electronics Engineering, Gebze Technical University, Kocaeli, Turkey, e-mail: okucur@gtu.edu.tr.}}

\maketitle

	\begin{abstract}
 Connected autonomous vehicles (CAV) constitute an important application of future-oriented traffic management. 
  A vehicular system dominated by fully autonomous vehicles 
  requires a robust and efficient vehicle-to-everything (V2X) infrastructure that will provide sturdy connection of vehicles in both short and long distances for a large number of devices, requiring high spectral efficiency (SE). 
 Power domain non-orthogonal multiple access (PD-NOMA) technique has the potential to provide the required  high SE levels. In this paper, a vehicular PD-NOMA testbed is implemented using software defined radio (SDR) nodes. 
 The main concerns and their corresponding solutions  arising from the implementation are highlighted. The bit error rates (BER) of vehicles with different channel conditions are measured  for mobile and stationary cases. The extent of the estimation errors on the success rate beyond the idealized theoretical analysis view  is investigated and the approaches to alleviate these errors are discussed. Finally, our perspective on possible PD-NOMA based CAV deployment scenarios is presented in terms of performance constraints and expectancy along with the overlooked open issues.
  
	
	\end{abstract}


\IEEEpeerreviewmaketitle

	\section{Introduction}

 \IEEEPARstart{C}{}onnected autonomous vehicles (CAV) is the technology in which autonomous vehicles communicate with other vehicles and infrastructures. Autonomous vehicles, stripped of human control, have ceased to be a public taboo and have become a technology that everyone is looking forward to. A more productive society with a decrease in accident numbers, reduced vehicle traffic and transportation time, improved vehicle-based energy efficiency (EE) and, consequently, new business models and scenarios in the industry are some of the benefits of CAV technology. To lead these developments, a novel model is required to be able to uplift autonomous vehicles system, allowing vehicles to join the live communication network around them, which will be referred to as vehicle-to-everything (V2X).


 For CAV-V2X technology, the necessity of an infrastructure that will implement these features has emerged in a short time. Most recently, application-based usage areas are categorized 5G Automotive Association (5GAA) \cite{A1}. Emergency braking, intersection management; vehicle maintenance and remote control, adaptation of the vehicle to the driver and cooperation with passengers, autonomous driving procedures, working principles of the road used by the authorities, traffic density information units and routing, vulnerable road user (VRU) protection, depending on the automation level of the vehicle (LoA), are some of the considered different use cases may be implemented or not. LoA is a classification method of advanced V2X applications, which indicates the ability and requirements of a system to be able to perform in an autonomous environment. There are 5 levels of automation from no automation to full automation; 
therefore, the use cases of each level may differ from the others, e.g. in order to share visual data with another vehicle in the same lane, a vehicle should satisfy 15 Mbps data rate, max. 50 ms latency and $99\%$ reliability within 80 meters range.


\setstcolor{blue} Today, orthogonal multiple access (OMA) based systems providing services to a single user in the same time/frequency block restrict this demand. Yet, OMA resources had long been exhausted. Considering that urban and rural environments will join the cellular networks, what is called cellular vehicle-to-vehicle (C-V2X), along with the vehicles; spectral efficiency (SE) plays a fatal role for an exponentially increasing number of users.  \setstcolor{black}



\begin{figure*}[t!]
    \centering
    \includegraphics[width=0.90\textwidth]{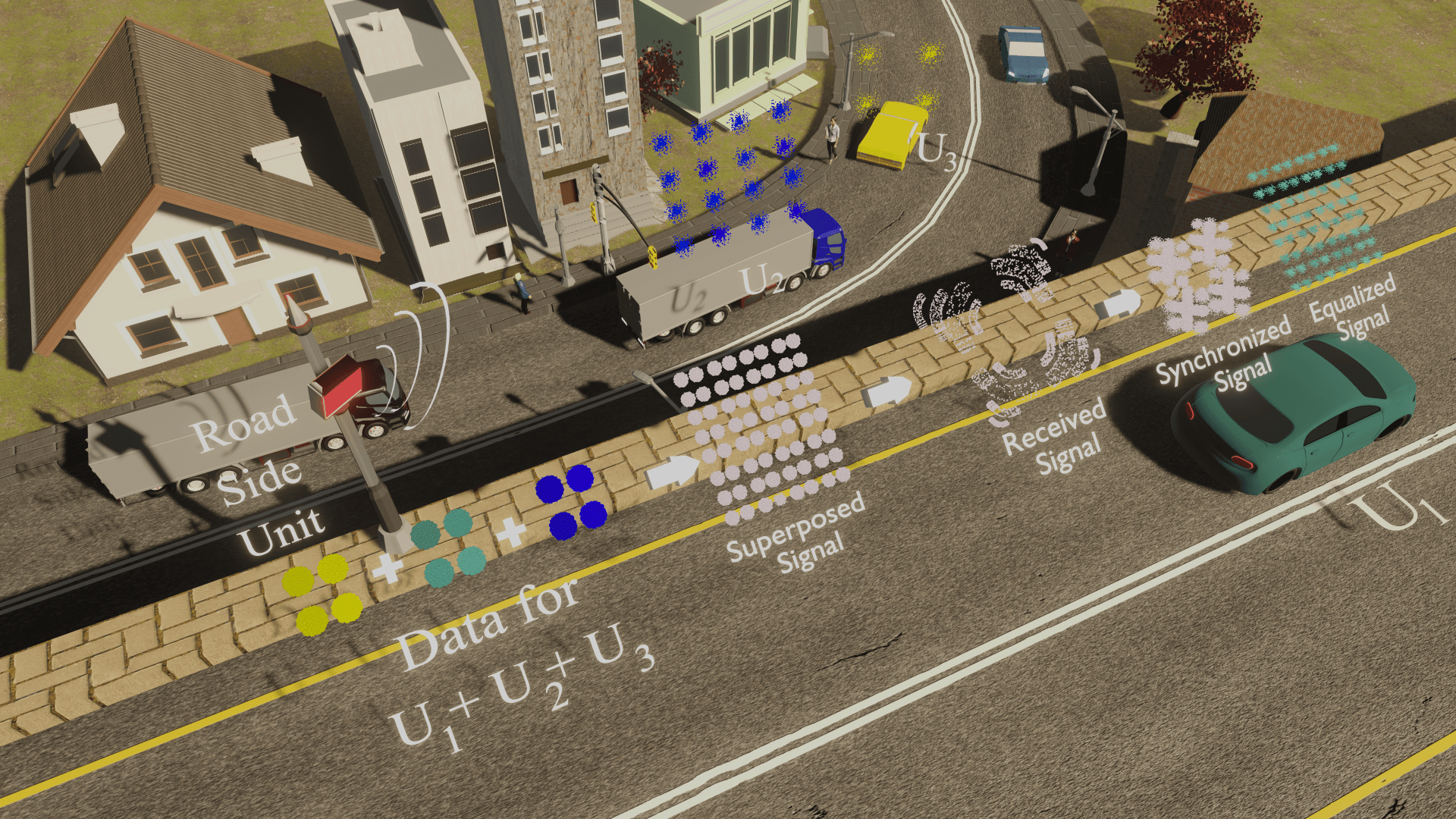}
    \captionsetup{justification=centering}
\caption{NOMA-based V2X deployment scenario including different CAV components, namely autonomous cars and vehicle platooning.}
    \label{fig:deployment}
\end{figure*}

The power domain NOMA (PD-NOMA) method, which distributes power between users from a single source, is examined in this study to increase SE over mobile vehicle units to clarify its suspects of its practical usage on possible CAV-V2X technology.

\subsection{A Cutting-Edge NOMA Scenario for CAV-V2X}
 \setstcolor{blue}
Non-orthogonal multiple access (NOMA) technique has some unique features for CAV-V2X systems, as it is flexible in that it can be developed just like OMA without requiring any preconditions and it provides user fairness while maximizing EE thanks to its power sharing. Fig. \ref{fig:deployment} illustrates the working principle of NOMA based communication network in a connected autonomous vehicular traffic based V2X system. Road side unit (RSU) transfers data of vehicles belonging to different channel structures. There are various deployments with different CAV components including autonomous cars and platooning trucks are shown in a sense of concept map. This concept has been taken as an example to our experimental 3-Users PD-NOMA test scenario.

\setstcolor{black}
 
In the context of standardization of NOMA and C-V2X, there are different outstanding efforts, especially in 3rd Generation Partnership Project (3GPP) \cite{XX}. 
Due to a strong interest in both academia and industry, the NOMA study is revived in Rel-15. After the completion of the NOMA work item, a study of implementing NOMA upstream transmission was carried out for 3GPP Release 17. It has been decided to continue NOMA based work items for the next releases beyond 5G networks. 

New Radio for vehicle-to-everything (NR-V2X) supports complex use cases including vehicle platooning, extended sensors, advanced driving for autonomous driving, and remote driving.

\setstcolor{blue}
\subsection{Motivation and Contribution} It seems from the literature review that there is no study with a real time implementation on the vehicular field of NOMA. With this research, we focus on examining the applicability of NOMA method in vehicular communications using the modulation and waveform techniques in use and the previously undiscovered problems it poses. Subsequently, feasible approaches and solutions applied to these problems are provided and demonstrated in the test environment. Finally, problems that have not yet taken place in literature and are planned to be studied are mentioned. In addition, we identify the advantages and limitations that NOMA brings to V2X by obtaining error performance measurements. The practical easinesses provided by software defined radios (SDR) in implementation and the problems they cause were given in detail with their key points. To clarify the novelty of this study, an achievement by empirical approach in NOMA powered C-V2X is gained by 3-Users case and this approach revealed the gaps that could not have been seen with theoretical analyzes.
\setstcolor{black}

The rest of the paper is organized as follows. The fundamental concepts for NOMA and CAV with standardization efforts are provided in Section II. The details of practical PD-NOMA implementation in V2X utilizing SDR nodes are presented in Section III. In Section IV, we discuss the open issues related to the different aspects of V2X systems that need to be considered in real-time deployment scenarios. The paper is concluded in Section V.

\section{Fundamental Concepts}

\subsection{Connected Autonomous Vehicles}
\vspace{1mm}
NOMA creates a new dimension for vehicle communication with its high efficiency, massive connectivity and low latency features it allows \cite{bagheri20205g}. 
One of the recent NOMA integrated V2X studies \cite{6} mentions a new dynamic resource allocation structure and scheduling process compatible with the successive interference cancellation (SIC) process for low-latency and high-reliability (LLHR) systems. As a forward-looking study of NOMA, \cite{7} discusses the principles of multiple-input/multiple-output (MIMO) NOMA in vehicular environments. In the realm of CAV studies; some of the latest advancements in CAV technology are discussed in \cite{B1} including vehicular network safety, reliability of vision sensing techniques and intersection management. In a real-time experimental study Long Term Evolution (LTE) for vehicle-to-vehicle (V2V) and Wi-Fi for vehicle-to-pedestrian (V2P) communication schemes are implemented to measure end-to-end latency and delivery rate as a dedicated short-range communications (DSRC) application \cite{C1}. Another practical study realizes vehicular visible light communication with higher order modulation techniques using SDRs \cite{D1}. 
\setstcolor{blue}
As in SDR deployment, \cite{moriyama2019experimental} evaluated uplink (UL) stationary NOMA with 5 users to imitate an Internet of things (IoT) system deployment. In addition to that, another recent study  shows that an SIC technique associated with channel state information (CSI) and quality of service (QoS) has the ability to remove limitations in outage probability \cite{15}. 
On the other hand, there are real-time implementation studies of the NOMA technique by utilizing SDR nodes \cite{2,3}. In \cite{2}, a WiFi prototype focused on the proposed frame structure in the physical layer is tested while conducting performance analysis of NOMA system based on bit error rate (BER) in \cite{3}.
\setstcolor{blue}

One of the recent proof-of-concept research \cite{cisco} belongs to Cisco and Verizon with mobile edge technology supported C-V2X deployment in 2022. This demo test tries to minimize RSU dependency with modular integrated routers. While it has been shown that low latency criteria for C-V2X can be met by lower cost routers, average vehicle-to-infrastructure (V2I) end-to-end latency was measured as 42.833 ms.
Verizon and Nissan North America’s Research and Advanced Engineering have performed one another C-V2X proof-of-concept study to test the sensor networks over vehicle devices in some use cases such as unpredictable pedestrian scenarios and unprotected vehicle turns in 2021. Cooperative associations and joint projects regarding C-V2X deployment among the large automotive industry have been discussed in \cite{field}.

\setstcolor{black}
\subsection{NOMA}
\setstcolor{blue}
\setstcolor{black}

It was previously mentioned that NOMA is a unique oppurtunity in order to be able to control user density with a more advanced multiple access method by providing low latency and reliability conditions with massive connectivity \cite{aldababsa2018tutorial}. 
Since the NOMA technique does not depend on orthogonality compared to OMA solutions such as code-division multiple access (CDMA) and orthogonal frequency-division multiple access (OFDMA), it can partially get rid of resource scheduling, planning operations and emerges as a candidate approach for expected low latency applications.
Thus, NOMA can address not only massive connectivity issues in ultra-dense networks (UDNs), but also the inevitable delay in ultra reliable low latency communications (uRLLC) applications. 

\setstcolor{blue}
In this study, we focus on examining the effect of PD-NOMA on vehicular connectivity.
The applications of PD-NOMA in downlink (DL) and  UL channels are differentiated from each other. It is realized by transmitting the superpositioned signal of the user signals at the base station (BS) in DL channels; while it is formed by combining the user signals with the corresponding power coefficients determined by the users or BS in the UL transmission. 
Although other users are perceived as noise for the weak user (User-1), strong users (User-2 and others) must use a multi-user detection (MUD) method in order to extract their data from the composed signal. 

Users apply SIC to the superimposed signal they receive as much as the weaker user signal while removing their own signal.
\setstcolor{black}

\section{A Practical NOMA Implementation in V2X}
\vspace{1mm}

\subsubsection*{Test Environment and Process}
\begin{figure}[t!]
    \centering
    \includegraphics[width=0.5\textwidth]{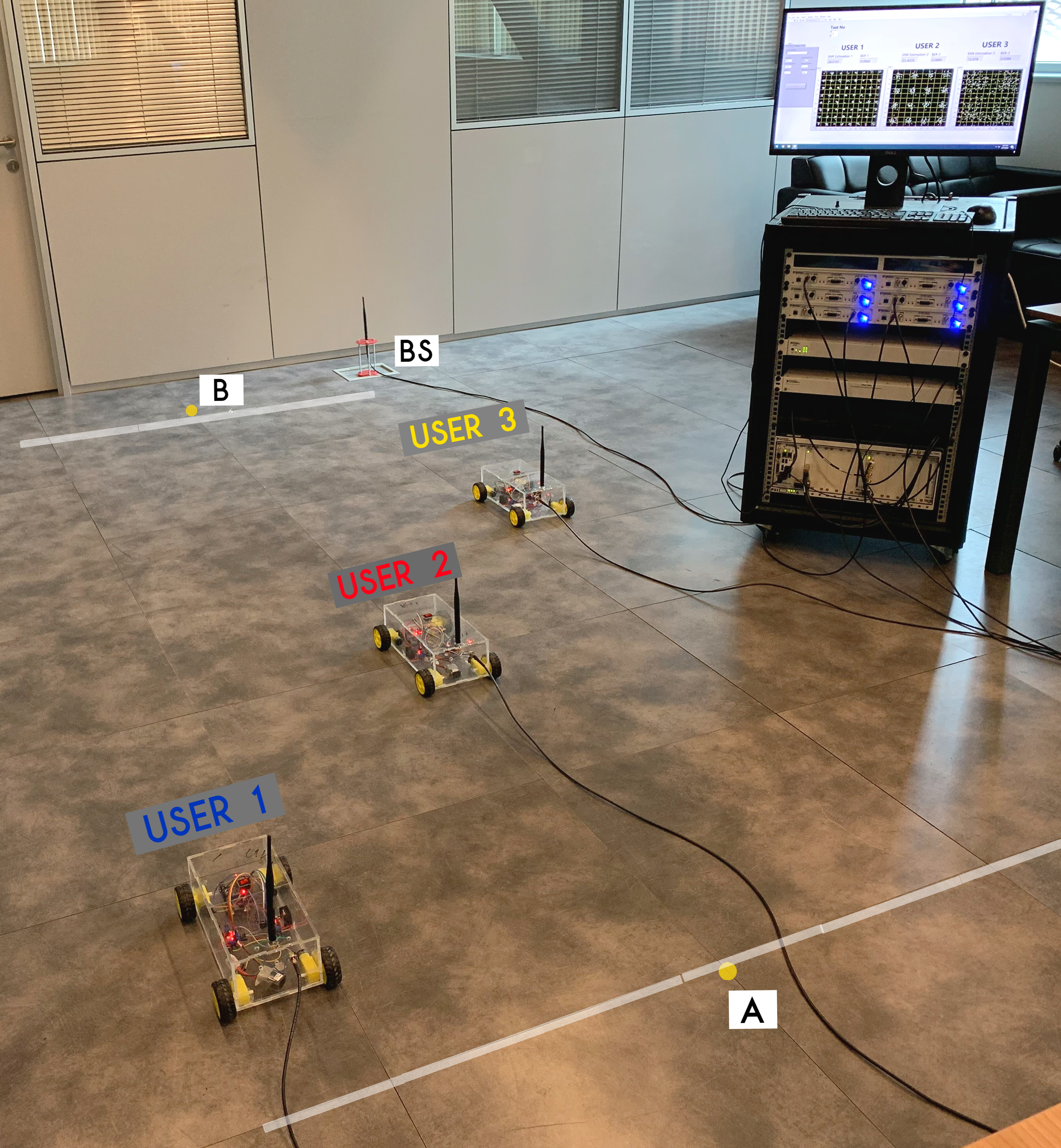}
   \caption{A snapshot of the 3-Users NOMA DL tests including start (A) and finish (B) lines.}
    \label{testbed}
\end{figure}

The testbed configuration consisting of three mini vehicles designed to represent the CAV system is shown in Fig. \ref{testbed}. Each vehicle has an Arduino microcontroller to operate, two L298N motor drivers for speed-direction control, and four direct current motors to move vehicles in a plain path. The shortest distance between lines A and B was determined to be 360 cm. Meantime, the vehicle farther from the BS has been named as User-1, the middle one as User-2, and the closest vehicle to the BS has been named as User-3. From the moment they depart, all 3 vehicles halt by reaching the B line after 3.63 s. Thus, the test that takes 5.74 s is terminated at the moment when vehicles stop.   

Multiple access data transfer from a BS to vehicles is provided by USRP model SDRs. There are a total of 4 USRP-2943R, 1 CPS-8910 model switch that provides power and a radio-host connection, and 1 PXIe-1085 model host computer in the DL setup, which consists of 3 vehicles (receivers) and 1 transmitter as BS. USRP-2943R has instant 40 MHz real time bandwidth with 16 bit digital to analog converter and 14 bit analog to digital converter resolution. The radios have a maximum output power of 20 dBm and input power of -15 dBm  at the operating frequency of the tests. It should also be noted that the devices expose 5 to 7 dB noise figure that needs to be taken into account for low power measurements. 

Test measurements were performed using 4-quadrature amplitude modulated (4QAM) multi-carrier orthogonal frequency-division multiplexing (OFDM) waveform with carrier frequency of 2.34 GHz and a bandwidth of 800 kHz. Before start, while the closest distance from the User-3 to the BS is 390 cm, the distance from the User-1 to BS is 427 cm. Meanwhile, the distance among vehicles is kept equal therefore, it can be assumed that the users' channel gain order has not been changed. Thereby, the burden of continuously assigning different power to users is been relieved. In a real-life scenario, as long as the channel coherence time is maintained, the vehicles’ CSI will be preserved and power distribution can be made accordingly. All parameters used in the tests are given in Table \ref{table}.

Signal processing on the transmitter starts with the generation of random data for each user. 1250 bits of data are converted into 625 symbols with 4QAM modulation operations and aligned into 5 OFDM symbol vectors as each of them contains 125 symbols. Interleaving pilot symbols into symbol vectors results in a total of 150 subcarriers for each OFDM symbol. Upon this, transmission signal operations continue with the zero padding before inverse fast Fourier transform (IFFT) operation. For the sake of synchronization and multipath fading prevention in the receiver, 64 symbol length of cyclic prefix is prepended to each vector. Eventually, the operations on the BS are completed and the signal is transferred to the users.
Unlike stationary systems, the Doppler shift and signal-to-noise ratio (SNR) estimation errors caused by the velocity of vehicles become a dominant contributor to carrier frequency offset (CFO), leading to intercarrier interference. In order to examine this problem, the Doppler frequency shift in the operating frequency was calculated as 6 Hz, with the average speed (0.876 m/s) found with the running time (3580 ms) of the vehicles and the total distance (313 cm) they took. In the matter of synchronization, for all vehicles, cyclic prefix based joint maximum likelihood (ML) estimator of time and frequency offset was detected and corrected at the receiver without additional pilots. Following synchronization and fast Fourier transform (FFT) conversion, channel estimation was performed with the least square (LS) method using the linear regression of the pilots added into the data; then, the channel compensation process was performed using the channel coefficients with the zero-forcing equalizer. By measuring error vector magnitude (EVM) values of the pilot signals, SNR estimation is made for each user. This process is followed by SIC for the User-2 and User-3 as mentioned in Section II. After each vehicle obtained its own data, the test is terminated after BER analysis.

As shown in Figure \ref{testbed}, the testing process takes place in two consecutive stages, stationary and mobile. After the start of the data flow with the stopwatch, the 3 cars, which remain motionless for exactly 2.165 seconds for the stationary stage, take off from A to B lane, maintaining the distances among them with equal and constant speeds. During the test; SNR, BER and CFO measurements of each vehicle are taken to analyze their performances. In addition to this test, the distribution of 6.2 million absolute channel coefficients was examined in order to estimate the channel distribution formation of the indoor environment. The non centrality and scale parameters were obtained with the ML estimation and the Rician K-factor was found to be 10.92. 

\subsection{Key Considerations Throughout the Measurements}

\vspace{1mm}
\subsubsection*{SNR estimation and outage measurement uncertainties within weak user}
There are some critical and noteworthy issues encountered during test taking. As one of them, it is not possible to calculate the SNR value of the superimposed signal using the data due to the SIC operations that must be performed before the demodulation of the signals in the DL. The SNR value to be obtained after the SIC process may not reflect the truth due to SIC imperfection. Channel and CFO estimation errors also contribute to exacerbating this error. In short, there may be further inaccuracy than usual in the SNR estimations for NOMA networks. What's more, users with multiple stage SIC tend to lose more data than other users. Another drawback of SNR estimation, errors appear in outage analysis. Any kind of misdetected SNR values may lead to an outage in users. In order to solve this problem, instead of using the EVM  calculation of the users' data, it is more meaningful to estimate the SNR value with pilot or training symbol assisted power calculation. Since the use of repetitive training symbols will decrease SE, the most reasonable SNR estimation method in NOMA-based data transfer is the utilization of sufficient number of regularly interleaved pilots interleaved with the data.

As a result of the same problem, outages may not be tracked flawlessly due to the failure in possessing threshold level reasoned by the error in the pilot or barker-based SNR estimations. An outage analysis requires an SNR estimation method with surgical precision since it is directly associated with the SNR value. With a similar approach, having a high rate of pilot/data would reach higher precision of instantaneous SNR values. One another approach could be using conditional flexible channel based estimators. Different noise levels in different channels may evoke variable kinds of estimators such as ML estimator which makes use of the spectral shape of the received signal or signal-to-variation ratio estimator.

\begin{figure}[t]
    \centering
    \includegraphics[width=0.495\textwidth]{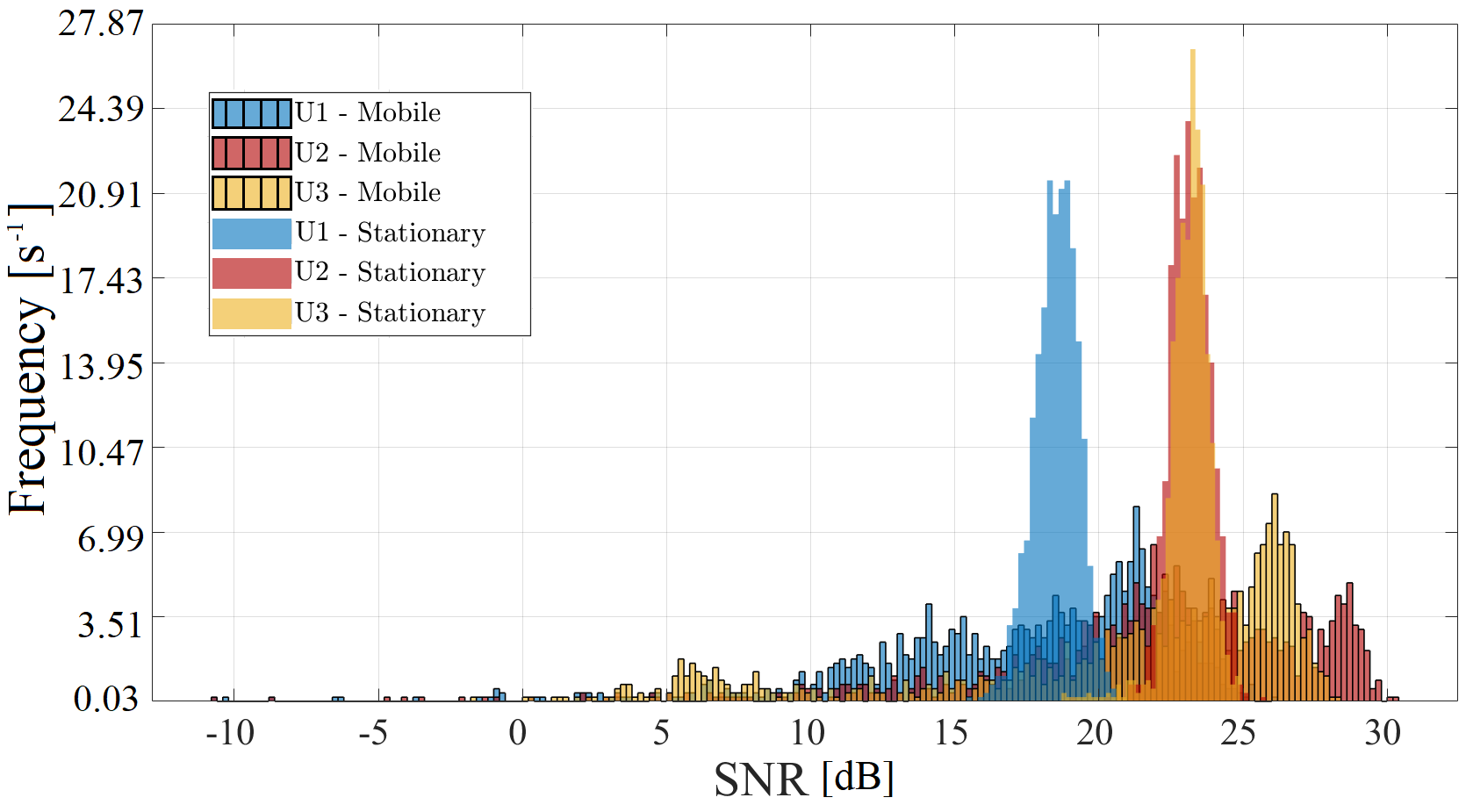}
    \caption{An examination of SNR occurrence frequency in stationary and mobile stages of tests.}
    \label{histogram}
\end{figure}

Histograms of the SNR measurements taken during the tests were shown in Figure \ref{histogram} and the SNR distributions of moving and stationary cases were compared. Since there is no practical way to calculate the exact SNR value of the received signal, it is quite logical to verify itself with such comparisons whose behavior can be predicted.
While the SNR values of the stationary vehicles at the beginning of the test create a chi-square distribution, the variance increases with the increase in the amount of instantaneous SNR deviation of the moving vehicles.  For example, the second user receives an average of 24.33 times 23 dB SNR per second in stationary tests, while an average of 2.47 times 23 dB SNR per second in mobile stage. While each user’s SNR distribution can be expressed as Gaussian when stationary, they cannot be modeled with a single function in the mobile position. Furthermore, it can be also interpreted that the increase in the average SNR of mobile vehicles in the tests is due to their decreasing distance from the BS.

A solid solution to SNR degradation issues caused by CFO and long-distance coverage issues is phased array antennas or beamforming the superimposed signal at UL and users in DL. Multi-beam antennas and phased arrays not only increase the directivity for the interested vehicle, but also decrease the inter-user interference among full duplex NOMA users in UL transmission. The beamforming technique not only prevents passive eavesdropping in public areas but also provides information security for the NOMA UEs that share both time and frequency resources without a randomization algorithm. One another benefit of beamforming in NOMA in terms of security is that a private user (or cluster) can benefit from improved signal-to-interference-plus-noise ratio (SINR) to increase signal quality and avoid any interference sourced physical attack methods such as jamming.

\subsubsection*{Causes and consequences of vulnerable synchronization}
\setstcolor{blue}
Unknown environmental factors contribute to the system's CFO amount, as much as hardware impairments. As the CFO prediction error of mobile systems increases, the compensation will be incorrect and as a result, the receiver gets a corrupted data. To make matters worse SNR estimation value will also be incorrect. In short, one estimation error will blunt the other which is more evident in NOMA users. For the strong users of the NOMA system performing SIC, the error magnitude increases cumulatively if they apply the SIC process using partially erroneous data whose offset has not been corrected.
\setstcolor{black}
In Figure 4, the instantaneous SNR values of the OFDM symbols, depending on time, are shown. With the movement that started in the 2.11th second, the increase in the instantaneous SNR changes can be seen. Together with the prediction errors, this increase becomes difficult to control. The same figure also gives the instantaneous BER and CFO values depending on the time from end to end. It seems obvious that while the estimated CFO values are stable of stationary vehicles belonging to the first half of the test, dramatic ups and downs begin with the movement. Additionally in Figure 5, BER of the same test has been obtained. While the BER values are calculated with the ratio of the number of erroneous bits to the total number of bits, the BER value is assigned to 1 for corrupted and undetected data arrays which had no place in BER calculations. Positive BER values start to show up right after the start off and during the mobile stage. It was observed that while the BER of the stationary vehicles were 10$^{-3}$ and below, the BER values of the vehicles in motion could rise up to 10$^{-1}$ and above time to time.

\begin{figure}[t!]
    \centering
    \includegraphics[width=0.5\textwidth]{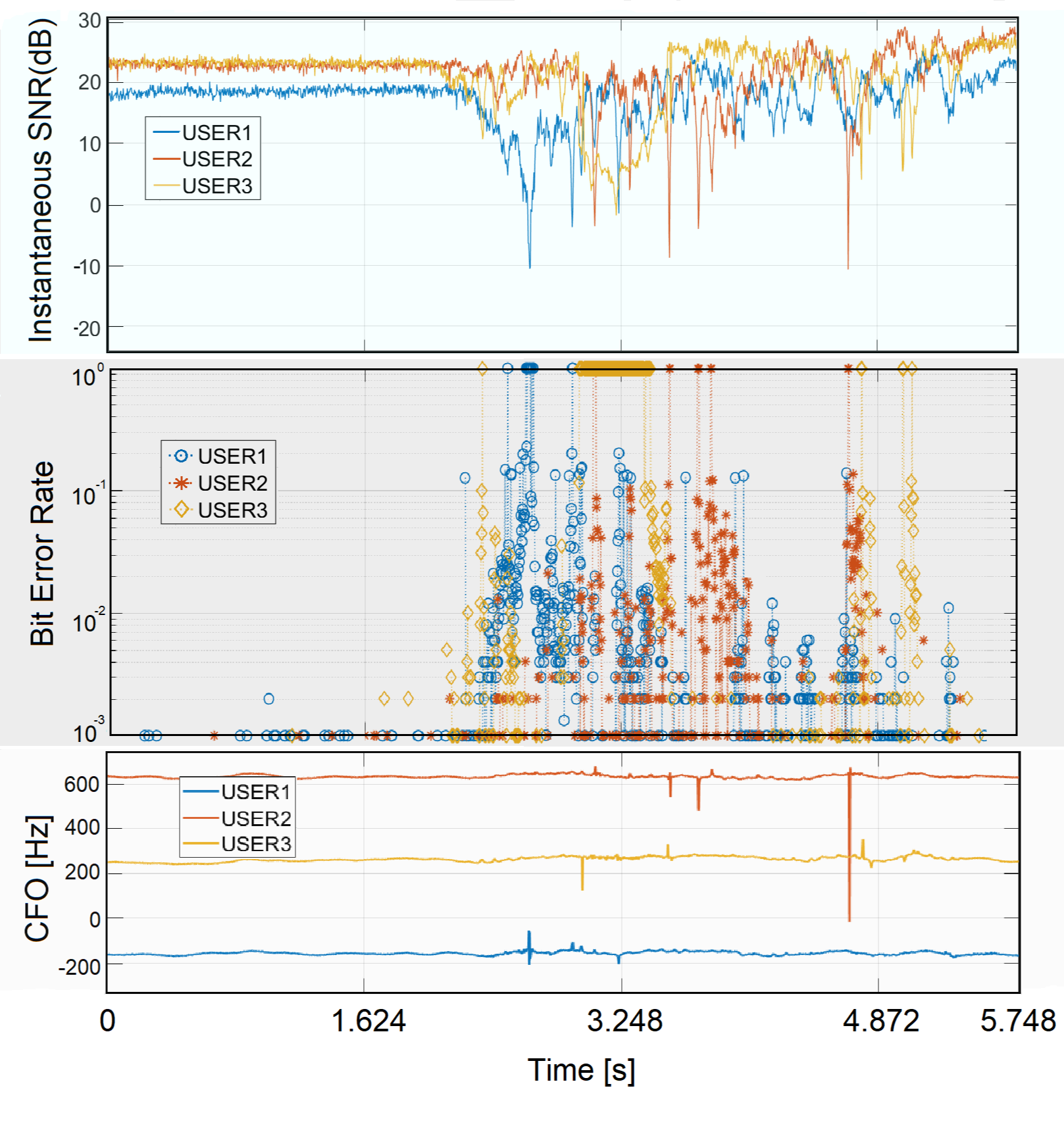}
    \caption{Fluctuation of signal performances over time.}
    \label{SNR}
\end{figure}




As mentioned earlier, the SNR value is calculated right after CFO estimation and compensation. This means that the measurement error in the CFO is also reflected in the SNR values. It is also seen in the test results that the dramatic changes in SNR and CFO overlap with each other. Thereupon, SNR decrease and BER increase can be seen in the peak values of the CFO.


\vspace{1mm}
\subsubsection*{Noise issues for higher order modulation}

In higher modulation techniques, the effect of noise over the incoming signal becomes more prominent. As a result, SIC errors will also be higher than usual. The error performance, which gets weaker with the increase of modulation order, becomes even weaker in superpositioned NOMA signals. However, this problem can be solved with a far apart selection of power coefficients, high precision channel estimation with frequent pilots and proper synchronization processes. It should be noted that strong users who will be applied SIC are the ones who will be most affected by this error increment caused by higher modulation levels. Briefly, the more SIC layers users have, the more errors they are exposed to. Although this problem can be avoided by increasing the power share of strong users, this solution will drag the system into other problems in vain such as performance loss in weak users.


\begin{table}[tb]
	\centering
	\caption{System Parameters}
	\label{table}
	\resizebox{0.48\textwidth}{!}{%
		\begin{tabular}{@{}l|c|c@{}}
			\toprule\midrule
			\multicolumn{2}{c}{Parameters} & Values \\  \hline \midrule
			\multicolumn{2}{l|}{Modulation}    & $4$QAM\\ \hline
			\multicolumn{2}{l|}{Carrier Frequency}    & $2.34$ GHz  \\ \hline
			\multicolumn{2}{l|}{Bandwidth}   & $800$ kHz\\ \hline
			\multicolumn{2}{l|}{I/Q Data Rate}    & $500$ kS/sec  \\ \hline  
			\multicolumn{2}{l|}{Power Coefficients (far to near)}   & $0.761, \, 0.191,\, 0.048$\\ \hline
			\multicolumn{2}{l|}{Transmitter Gain}   & $15$ dB \\ \hline
			\multicolumn{2}{l|}{Receiver Gain}   & $10$ dB \\ \hline
			\multicolumn{2}{l|}{Average Speed}   & $0.876$ m/s \\ \hline
			\multirow{3}{*}{Distance to BS at Starting point (A)} & User-1 & $4.27$ m \\ 	&User-2 & $4.02$ m  \\ & User-3 & $3.9$ m \\ \hline 
				\multirow{3}{*}{Distance to BS at End point (B)} & User-1 & $1.25$ m \\ 	&User-2  & $1.12$ m \\ & User-3 & $0.57$ m \\
				\midrule\bottomrule
				
		\end{tabular}%
}
\end{table}

\subsubsection*{Interference among users in DL}

One of the challenges that comes with PD-NOMA is that users sharing the same power interfere with each other. The more users system operates, the more users affect each other. This effect is characterized by determining factors such as the number of users sharing the power, the distance between users, power coefficients and carrier frequency; it is possible to minimize this problem with both practical and analytical approaches. 

To prevent self-interference for multi-user SDR applications the listed guidelines are followed: 
\begin{enumerate}
    \item The appropriate power coefficient values are estimated.
\item In the measurements tested at distant power coefficient values, to be able to reduce the effect of the noise on the system, the transmitter and receiver output power were kept constant, while the received power by the users were become variant due to the changing distance. Thus, the number of hardware changes affecting the system has been reduced to one which is source transmitter radio.
\item Interference mitigation techniques for dense topologies are  used as matching strong and weak users by channel gain difference. 
For this reason, analyzing users in terms of error performance or bit rate, SINR values of users should be taken into account.
\setstcolor{blue}
\item Due to the nature of NOMA, the synchronization of the carrier signal becomes deadly critical. Well synchronized users using the same timing reference will decrease the user interference dramatically. Likewise, nonlinear frequency and phase offset estimation methods are costly but rather solid options. 
\end{enumerate}

\subsubsection*{Power allocation effect in dynamic channels}
\setstcolor{blue}
Depending on the line of sight, channel difficulty and vehicle speeds, the power coefficient values assigned to the users may not be effective at all times. This reveals the necessity of CSI feedback of each user to source. A power distribution algorithm is required to keep user fairness at the maximum level. As a solution to this problem, a power allocation method was formulated so that power coefficients proportional to the square of the BS distance were assigned to each user. It leads to keep the channel impact level minimal. Thus, not only the weak user's connection will be compensated, also the interference between users will be minimized due to the increased power difference.

\begin{figure}[t]
    \centering
    \includegraphics[width=0.48\textwidth]{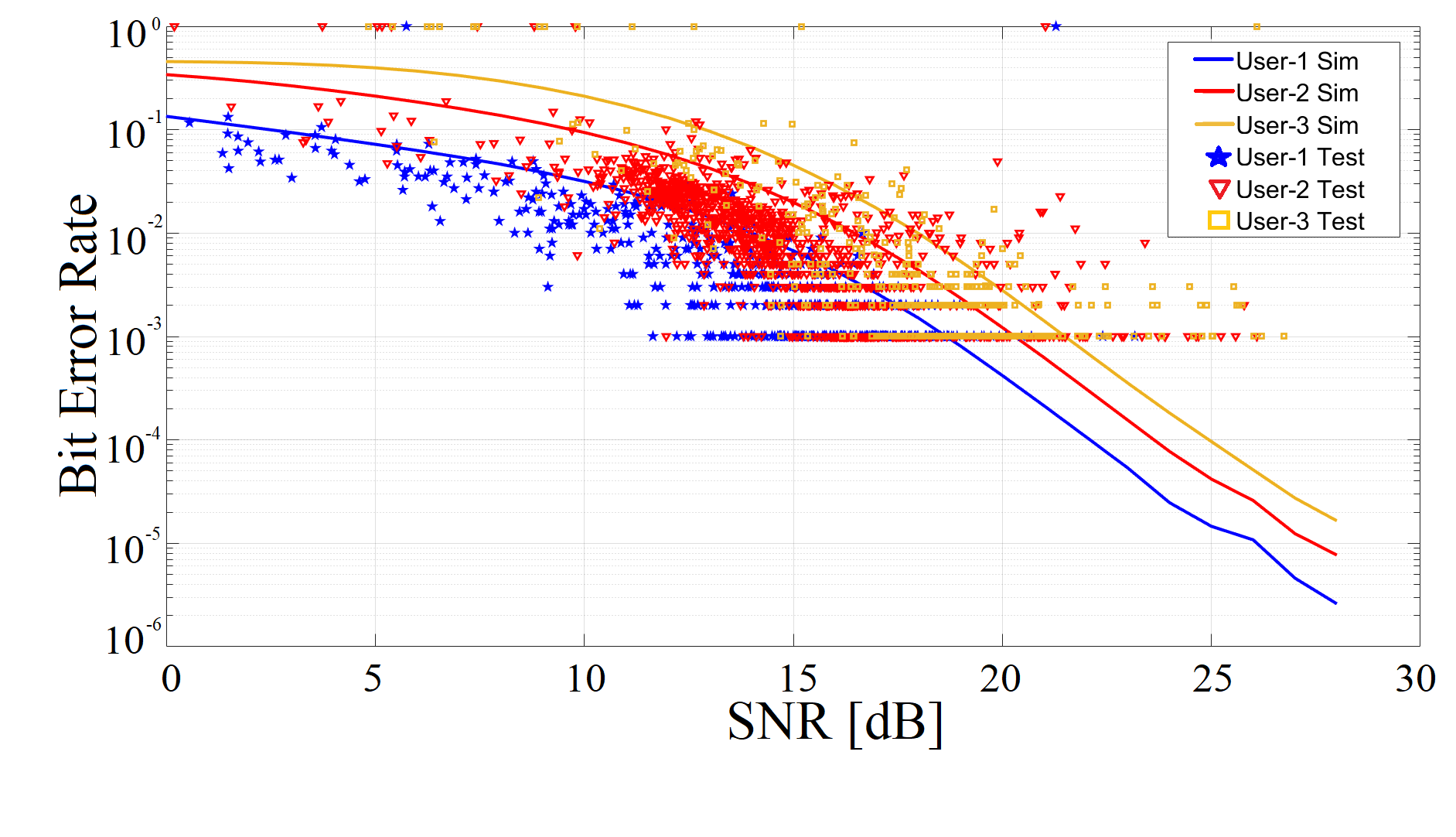}
    \caption{Performance comparison of mobile NOMA vehicles.}
    \label{sim}
\end{figure}

For this purpose, in order to see how accurately the power coefficients can be realized in real time experiments, the samples obtained during the mobile stage of the V2X tests were compared with the 3-Users PD-NOMA simulations. Note that the Rician channel factor has been found and used in simulation as 10.92 as the result of a realtime test. In Figure \ref{sim}, the results of this comparison are shown in terms of BER performances at different SNR values.
It seems that with the power parameters used, all users possess an error floor with bit error performance in the order of one thousandth. The matching results with the tests prove that the PD-NOMA system is indeed realizable.

\section{Open Issues}

\paragraph*{Regulations, safety, security and privacy}

The standardization activities that also need to satisfy all the regulation requirements are critical in the domain of autonomous vehicles. The connectivity between vehicles needs to be robust and safe to provide reliable services. Physical layer related issues associated with rapidly varying power levels and high Doppler shift need to be properly addressed for robust operating service levels. One of the anti-jamming methods is spectrum scan which is a quick and efficient problem solver that gives the link flexibility to find out a convenient band before and during jamming attacks. One another way is that the beamforming the transmission that results with high directivity would boost the SINR of the reception which can suppress the jammer attacks. Likewise, cognitive radio technology has the ability to escape to free channels without interference which would prevent users to get affected by a signal collision. Furthermore, the security and privacy aspects of NOMA users need to be investigated against threats from both the eavesdroppers or other users due to SIC process.

\paragraph*{Cognitive features and device-to-device connectivity}
In order to further improve the SE, cognitive features can be incorporated with NOMA signaling to determine the suitable transmission bands. To further improve the SE per area, coordinated transformation can also be coupled with device to device connectivity when NOMA users located at proximity aim to transmit to one another. Such features can also aim to improve the EE of the system; however, the operational performance expectations must be properly monitored with safety issues in mind. Furthermore, the target number of users' power coefficients and the resources for transmission bands can be determined by the use of machine learning principles. 

\paragraph*{Limitation of Use Cases}

A major flaw of the merging technology of C-V2X with NOMA is that system complexity that limits possible use cases. NOMA increases system efficiency in many ways while loading some challenges into users as a burden, such as error lower bound, user scheduling complexity, dependency on CSI feedback, additional CFO caused by the Doppler effect, etc. With the removal of these flaws, a wide range of research area with full of opportunities will emerge. DL NOMA and C-V2X systems jointly add complexity to users because of user detection and sensor information processing. For this reason, very large scaled integration (VLSI) applications that will increase computational performance have the potential to be the only solution to this problem. In a similar concept, CSI feedback to the transmitter in mobile NOMA systems could solve multiple issues at the same time by allowing transmitter beamforming, precoding and finding the favorable propagation for MIMO systems.
\setstcolor{black}
\paragraph*{Integration of principles of software defined networking}
The principle of separating the control plane from the data plane can also be coupled closely with the use of NOMA signaling. For instance, control signals can remain OMA transmission for improved reliability, yet the test signals can be transmitted over NOMA principles with the goal of improving the SE. In another scenario, control signals requiring a lower data rate compared to the data signals and low-latency can be considered as a weak user's signal, i.e. low-rate delay sensitive signals. 


\section{Conclusion}
In this study, the demands of CAV-V2X for a real time deployment has been discussed and it has been shown that a NOMA based CAV scenario is self-evident by experimental observation of real-time mobile SDRs. It has been elucidated that a robust multiple access method is required for autonomous vehicles, whose capabilities increase with their amount at the same time. 
We believe that NOMA is the renaissance to meager multiple access techniques of current vehicular networks; and in this manner, the demands of the incoming autonomous vehicles technology matches the capabilities of PD-NOMA utterly, as shown through experimental results.

\vspace{0.5cm}


\newpage
\ifCLASSOPTIONcaptionsoff
  \newpage
\fi

\bibliography{refs}
\bibliographystyle{IEEEtran}
\section*{Biographies}
\footnotesize{

ERAY GUVEN (guven.eray@polymtl.ca) received his B.S. degree in Electronics and Communication Engineering at Istanbul Technical University, Turkey in 2021. He is currently studying for a Ph.D. degree at  Polytechnique Montr\'eal, Montr\'eal, Canada

CANER GOZTEPE (goztepe@itu.edu.tr) received his B.S. degree in Electronics and Communication Engineering and MSc. degree in telecommunication engineering from Istanbul Technical University, Turkey in 2017 and 2019 respectively. He is currently studying for a Ph.D. degree at Istanbul Technical University.

MEHMET AKIF DURMAZ (durmazm@itu.edu.tr) received his B.S. degree in Electronics and Communication Engineering and MSc. degree in telecommunication engineering from Istanbul Technical University, Turkey in 2018 and 2021 respectively. He is currently working in Turkcell Technology as R\&D Engineer. 

SEMIHA TEDIK BASARAN (tedik@itu.edu.tr) received her B.S. degree, MSc. degree and PhD degree from Istanbul Technical University, Turkey. She is currently an assitant professor at the Department of Electronics and Communication Engineering, Istanbul Technical  University, Istanbul, Turkey. 

GUNES KARABULUT KURT ({gunes.kurt@polymtl.ca}) received the Ph.D. degree in Electrical Engineering from the University of Ottawa, Ottawa, ON, Canada, in 2006. {She is with the Department of Electrical Engineering,  Polytechnique Montr\'eal, Montr\'eal, Canada}. 

OGUZ KUCUR (okucur@gtu.edu.tr) received his Ph.D. degree in Electrical and Computer Engineering from the Illinois Institute of Technology (IIT), Chicago, USA in 1998. He is currently a full professor with the Department of Electronics Engineering at Gebze Technical University, Kocaeli, Turkey.}

\end{document}